# Structure–Behavior Coalescence and the Limits of Traditional Systems Theory

William S. Chao

**Structured Abstract**

**Purpose:** This paper examines a foundational assumption in traditional systems theory—namely, the separability of structure and behavior—and argues that it underlies persistent limitations in explaining system identity, emergence, and boundaries. The study introduces Structure–Behavior Coalescence (SBC) as a reframing principle addressing this issue.

**Design/Methodology/Approach:** A conceptual and theoretical analysis is conducted across structural, behavioral, and hybrid systems approaches. On this basis, the Structure–Behavior Separation Problem (SBSP) is formulated as a foundational constraint in systems theory. Structure–Behavior Coalescence (SBC) is then developed as a meta-theoretical reframing grounded in cybernetic and process-oriented systems traditions.

**Findings:** The analysis shows that treating structure and behavior as independently specifiable primitives leads to underdetermined system identity and incomplete explanations of emergence and boundary formation. SBC resolves this by conceptualizing structure and behavior as relationally interdependent aspects of a single systemic process, where system identity arises from their persistent co-determination.

**Originality/Value:** SBC provides a unified reframing of system constitution that replaces representational dualism with structural–behavioral coalescence. It offers a process-based account of system identity consistent with second-order cybernetics and complexity-oriented systems theory.

**Plain Language Summary**

This paper examines a foundational assumption in traditional systems theory, namely that structure (the organization of components) and behavior (the evolution of system activity over time) can be treated as separable analytical dimensions.

It argues that this separation contributes to persistent difficulties in explaining system identity, particularly in cases involving change, emergence, and boundary specification.

To address this issue, the paper introduces Structure–Behavior Coalescence (SBC) as a reframing principle. SBC proposes that structure and behavior should not be understood as independently existing entities that are subsequently related through modeling constructs, but as mutually constitutive aspects of a single systemic process.

From this perspective, system identity is understood as arising from the sustained co-determination of structural organization and behavioral dynamics, rather than from their external correspondence or alignment. This reframing provides a unified way of understanding system identity, emergence, and boundary formation within a cybernetically informed systems perspective.

## 1. Introduction

Systems theory has long aimed to provide a general framework for describing organized complexity across natural, technical, and social domains [1–2]. Despite its breadth and methodological maturity, a persistent limitation remains: systems lack a fully consistent theoretical account of identity. This paper addresses this limitation by examining the foundational assumptions underlying how systems are conceptually constituted. Methodologically, the paper adopts a conceptual-analytical approach grounded in comparative systems theory and cybernetic foundations.

In current systems theory, systems are described through two dominant representational modes: structural representations (e.g., components, relations, and organizational configurations) and behavioral representations (e.g., state transitions, processes, and interactions) [3–4]. These modes are widely used across systems engineering and cybernetics, yet they are typically developed as parallel descriptive layers rather than as a unified theoretical construct.

This separation gives rise to an implicit dual-view assumption: structure and behavior are treated as distinct but complementary aspects of the same system. Structural accounts emphasize organization and composition, while behavioral accounts emphasize temporal evolution and interaction. Although this duality supports practical modeling effectiveness [3,5], it leaves unresolved a more fundamental question of system constitution: how structure and behavior jointly determine a single system identity.

Across structural, behavioral, and hybrid modeling traditions, this unresolved issue manifests as an underdetermination of system identity. Similar structures may support multiple behavioral regimes, and similar behaviors may be realized by different structural organizations. Hybrid approaches partially address this issue through coordination mechanisms, but these typically rely on external mappings between representations rather than explaining their constitutive unity [6–7]. Consequently, system identity is often assumed rather than theoretically grounded.

These observations indicate that the limitation is not methodological but conceptual in nature. It reflects a deeper assumption that structure and behavior are independently definable analytical primitives, connected only through interpretation or modeling alignment. This assumption enables multi-view modeling approaches but simultaneously prevents a unified account of system identity, emergence, and boundary formation.

From this perspective, persistent problems in systems science—including identity under change, emergence, and system boundary determination—can be interpreted as consequences of this underlying separability assumption [8–9]. When systems undergo structural or behavioral variation, existing frameworks struggle to explain continuity of identity or the origin of system-level properties.

This paper identifies this underlying assumption as the SBSP. The SBSP is not treated as a representational inconvenience but as a foundational constraint on traditional systems theory. The central question is therefore not how to better coordinate structure and behavior, but whether their assumed separability is conceptually justified.

To address this issue, the paper proposes Structure–Behavior Coalescence (SBC) as a reframing principle. SBC conceptualizes structure and behavior not as independently existing components linked by correspondence rules, but as mutually constitutive aspects of a single systemic process. In this view, system identity arises from the persistent co-determination of structural organization and behavioral dynamics, rather than from their external alignment.

The remainder of this paper is structured as follows. Section 2 examines the limitations of traditional systems theory in detail. Section 3 formulates the SBSP as a foundational constraint. Section 4 introduces Structure–Behavior Coalescence and develops its implications for system identity. Section 5 discusses broader theoretical consequences for systems theory. Section 6 concludes the paper.

## 2. Limitations of Traditional Systems Theory

This section synthesizes limitations across structural, behavioral, and hybrid systems approaches as a unified analytical finding. Despite methodological diversity, these approaches share a common assumption: systems can be analyzed through a separation of what they are composed of (structure) and how they evolve over time (behavior) [1,3,4]. This section consolidates the limitations of this assumption as a unified critique of system identity, emergence, and boundary specification.

Structural approaches define systems in terms of components, relations, and organizational configurations [1,5]. Systems are thus modeled as relatively stable patterns of organization, supporting analysis of architecture, connectivity, and decomposition in engineering and network contexts [5–7]. However, structural descriptions alone are insufficient to determine system identity under change. Equivalent structures may exhibit divergent behaviors depending on environmental coupling or interpretive context, indicating that structure is necessary but not sufficient for capturing system dynamics or persistence [3,8].

Behavioral approaches instead define systems through state evolution, processes, and interaction sequences [3,4]. This enables effective modeling of temporal dynamics but abstracts away the organizational constraints that condition those dynamics. As a result, similar behavioral trajectories may be realized by structurally distinct systems, limiting behavior as a standalone criterion for system identity. Behavior, while expressive of change, does not uniquely determine the system that produces it.

Hybrid frameworks attempt to reconcile these perspectives by combining structural and behavioral views within layered architectures or multi-view modeling approaches [5,7,11]. However, the relationship between structure and behavior is typically specified externally through mappings, consistency constraints, or alignment rules rather than derived from a constitutive principle. Integration is therefore representational rather than ontological, introducing coordination complexity without resolving the underlying identity gap.

Across these approaches, persistent difficulties emerge in system identity, emergence, and boundary definition. System identity becomes underdetermined when structural and behavioral variations are partially independent [1,8,12]. Emergent properties are often described as arising from their interaction, yet remain weakly grounded in constitutive theory [8–10]. System boundaries likewise depend on modeling decisions rather than principled criteria [1,5]. These issues collectively indicate that traditional systems theory lacks a unified account of system constitution under change.

The common denominator of these limitations is the implicit separability of structure and behavior as independent analytical primitives [1–4,11]. Structure is treated as organizational substrate, and behavior as temporal unfolding over that substrate, with their linkage introduced through auxiliary modeling mechanisms [7,19–21]. This persistent separation suggests a deeper constraint: traditional systems theory does not explain why a particular structural–behavioral pairing constitutes a single system.

## 3. The Structure–Behavior Separation Problem

The limitations identified in the previous section indicate that difficulties in explaining system identity, emergence, and boundaries are not merely representational issues but reflect a deeper conceptual constraint [1,8–10,12–18]. This section formalizes the assumption underlying these limitations as a conceptual problem. This constraint can be formulated as an implicit assumption governing most systems-theoretic frameworks: the separability of structure and behavior.

In standard formulations, structure and behavior are treated as analytically independent primitives [1–5,13,15–18]. Structure is typically defined as the organization of components and relations constituting a system's composition, while behavior refers to the temporal evolution of states, processes, or interactions [3,4,6,11,19]. Although often described as complementary, they are rarely treated as constitutively interdependent. Instead, structure is assumed to exist as a static or quasi-static configuration, while behavior unfolds over or within it. Their relationship is then introduced through auxiliary modeling mechanisms rather than derived from a unified principle [7,19–21].

This assumption implies that structure and behavior are independently specifiable aspects of a system [13,15–18]. Under this view, one can describe structure without behavior and behavior without structure, subsequently linking them through mappings, alignment rules, or consistency constraints. This separability underpins common modeling practices such as multi-view representations and layered architectural decomposition [7,20,21].

However, this conceptual independence generates a deeper theoretical tension. If structure and behavior are independently definable, it becomes unclear what grounds their unity as a single system. The issue is no longer representational adequacy but constitutive explanation: why a given pairing of structure and behavior should constitute one system rather than two correlated descriptions.

This tension becomes explicit in system identity. Because structure and behavior may vary independently, identity becomes underdetermined. A single structural configuration may support multiple behavioral regimes depending on context, while similar behaviors may arise from distinct structural organizations [8–10,12,17,22]. Consequently, neither structure nor behavior alone provides sufficient criteria for system identity, and their combination does not inherently resolve this ambiguity.

A similar limitation appears in how structure–behavior relations are handled in existing frameworks. Their connection is typically established through external mechanisms such as mappings, transformation rules, or consistency conditions [7,19–21]. While useful for coordination, these mechanisms operate at the representational level and do not explain why structural and behavioral descriptions belong to a single system in a constitutive sense.

Taken together, these observations indicate that traditional systems theory lacks a principled account of how structure and behavior jointly constitute system unity [13–18,23]. This absence is not a gap within specific formalisms but a consequence of the deeper assumption of separability. The SBSP can therefore be stated as follows: systems theory presupposes that structure and behavior are independently definable, yet provides no constitutive principle explaining how their combination yields a coherent system identity. This formulation identifies the SBSP as a foundational constraint within traditional systems theory.

## 4. Structure–Behavior Coalescence

Structure–Behavior Coalescence, introduced as a response to the SBSP, is not proposed as a competing modeling formalism but as a reframing of the constitutive assumptions underlying systems theory. In continuity with second-order cybernetics, which emphasizes observer-dependence and the co-construction of systems and descriptions [33–34], SBC shifts attention from representational dualism toward relational co-determination. SBC is formulated as a theoretical reframing rather than a formal modeling framework.

The central claim of SBC is that structure and behavior are not independently specifiable primitives subsequently coupled through modeling operations, but mutually constitutive aspects of systemic organization. From this perspective, system identity does not precede the relation between structure and behavior but arises through their ongoing co-production. This aligns with process-oriented systems philosophies in which organization is defined dynamically rather than as static decomposition [23,25].

Within SBC, structure is reinterpreted as behavior-constraining organization: what is identified as structure is a stabilized pattern of constraints emerging from recurrent behavioral regularities. Conversely, behavior is reinterpreted as structure-realizing dynamics, through which systemic activity continuously reproduces and stabilizes the very organizational patterns later abstracted as structure. This is consistent with cybernetic notions of recursive causality and operational closure [3,4,32], where causation is distributed across ongoing system dynamics rather than assigned to fixed components.

System identity is therefore not grounded in correspondence between separately given structural and behavioral descriptions, but in the persistence of a coalescent relational invariant between them. Identity is a property of the stability of this co-determining loop, rather than of either structure or behavior in isolation. This shifts the explanatory focus from representation to persistence under recursive organization, consistent with Ashby's cybernetic emphasis on regulatory viability under perturbation [35].

Emergence is likewise reinterpreted. Rather than being an additive property arising from interactions between structural and behavioral layers, emergence is treated as an intrinsic feature of continuous structural–behavioral co-determination. System-level regularities are not imposed from above but continuously generated through distributed dynamics, consistent with complexity-theoretic accounts of non-decomposable systems [8–10,27].

Importantly, SBC does not eliminate structural or behavioral modeling practices. Instead, it repositions them as epistemic projections of a deeper coalescent process. Structural and behavioral models remain analytically useful, but their ontological status is revised: they are partial abstractions of a unified cybernetic organization rather than independent system dimensions [1,25,34].

At a meta-theoretical level, SBC is consistent with second-order cybernetics, where the distinction between system and observer is itself part of the system-theoretic domain [33–34]. It therefore reframes the core question of systems theory from how structure and behavior are related, to how their separability is conceptually produced and maintained.

## 5. Implications for Systems Theory

This section outlines five implications of SBC for systems-theoretic interpretation. It reframes how modeling practice, system identity, cybernetic regulation, and emergence are conceptually grounded.

A first implication concerns systems modeling. Classical systems engineering and multi-view modeling frameworks distinguish structural and behavioral representations as separate but coordinated views, typically related through mappings or architectural specifications [19–21]. Within SBC, this separation is not ontologically fundamental but epistemically derived. Modeling is therefore reinterpreted as the articulation of different observational projections of a single coalescent system rather than the integration of independently existing descriptions. This aligns with ISO 42010's view of viewpoints as partial representations rather than complete system accounts [21].

A second implication concerns system identity. Traditional approaches treat identity either as an invariant property preserved across change or as an emergent result of consistent alignment between structural and behavioral descriptions. SBC instead locates identity in the temporal persistence of structural–behavioral co-determination. Identity is thus processual and relational, consistent with process philosophy and cybernetic traditions that emphasize continuity through organization rather than static form [23,32].

A third implication concerns cybernetic feedback. In first-order cybernetics, feedback is typically modeled as an interaction between system components or between structure and behavior [3,4]. SBC instead treats feedback as an intrinsic property of a coalesced system, consistent with second-order cybernetics in which observer and system are embedded in a recursive loop [33–34]. Regulation is therefore not externally imposed but emerges from internal structural–behavioral coupling.

A fourth implication concerns emergence and complexity. Complexity theory often explains emergence through multi-level interactions or network effects [8–10,12,27]. SBC instead interprets emergence as arising from continuous structural–behavioral co-production within a unified process. This shifts emphasis from stratified levels to distributed causality and non-decomposability, aligning with relational accounts of complex systems [28–29].

Finally, SBC situates itself within broader relational and process-oriented traditions in systems theory [14–18,25–26]. SBC does not introduce a new representational layer but reframes existing ones. Many distinctions treated as foundational in traditional systems theory are therefore reinterpreted as methodological abstractions rather than ontological necessities.

## 6. Conclusion

This paper examined a foundational assumption in traditional systems theory: the separability of structure and behavior. Although structural, behavioral, and hybrid approaches have developed extensive modeling capabilities, they consistently rely on this implicit distinction, which continues to shape how system identity, emergence, and boundaries are conceptualized.

We formalized this assumption as the SBSP, arguing that it is not a methodological limitation but a deeper conceptual constraint embedded in systems-theoretic reasoning. In particular, the independent definability of structure and behavior generates an explanatory gap in accounting for system unity under change.

To address this limitation, we introduced Structure–Behavior Coalescence (SBC) as a reframing principle. SBC proposes that structure and behavior are not separable primitives but mutually constitutive aspects of a single systemic process. System identity, in this view, is grounded in the persistence of structural–behavioral co-determination rather than in either aspect independently.

The implications of this reframing extend across modeling practice, cybernetic interpretation, and complexity theory. Several long-standing issues—including identity persistence, emergence, and boundary formation—can be reinterpreted as consequences of the underlying assumption of separability rather than as independent theoretical problems.

Rather than replacing existing modeling frameworks or systems theories, SBC revises the interpretive assumptions through which structure and behavior are understood. It aligns systems theory more closely with second-order cybernetics and process-oriented traditions, where relational dynamics rather than static decomposition form the basis of system explanation. In this sense, SBC offers a reframing of system constitution rather than an additional modeling construct.